\documentclass{elsart}
\usepackage{amsmath}
\usepackage{graphicx}
\graphicspath{{figures/}}
\def\pmstretch{1.3}
\usepackage{ifthen}
\usepackage{color}
\include{diagrams}

\allowdisplaybreaks

\newcommand{\mathgraphics}[2]{
  \begin{minipage}[c]{#1}
    \includegraphics[width=#1]{#2}
  \end{minipage}
}
\newcommand{\MSbar}{\ensuremath{\overline{\text{MS}}}}

\begin{document}
\begin{frontmatter}
\begin{flushleft}
TTP11-25\\
SFB/CPP-11-48\\
arXiv:1110.5581
\end{flushleft}
\title{Low- and High-Energy Expansion of Heavy-Quark Correlators at
  Next-To-Next-To-Leading Order} 
\author[Turin]{A.~Maier}, 
\author[Karlsruhe]{P.~Marquard}
\address[Turin]{INFN, Sezione di Torino \& Dipartimento di Fisica
Teorica, Universit\`a di Torino, 10125 Torino, Italy}
\address[Karlsruhe]{Institut f\"ur Theoretische Teilchenphysik,
 Karlsruhe~Institute~of~Technology~(KIT), 76128 Karlsruhe, Germany}
\begin{abstract}
We calculate three-loop corrections to correlation functions of heavy-quark currents in the low- and
high-energy regions. We present 30 coefficients both in the low-energy
and the high-energy expansion of the scalar and the vector correlator with non-diagonal
flavour structure. In addition we compute 30 coefficients in
the high-energy expansion of the diagonal vector, axial-vector, scalar and
pseudo-scalar correlators. Possible applications of our new results are
improvements of lattice-based quark-mass determinations and the approximate
reconstruction of the full momentum dependence of the correlators.
\end{abstract}
\end{frontmatter}

\section{Introduction}
\label{sec:introduction}
The precise determination of the fundamental parameters of the Standard
Model (SM) is of utmost importance to pin down the fundamental
interactions of the elementary particles and to establish deviations
from the SM. Three especially interesting parameters, namely the strong
coupling constant and the masses of charm and bottom quarks, can be
determined with high accuracy from correlators of heavy-quark currents
by comparing { their} moments with experimental data using sum
rules~\cite{Shifman:1978bx,Novikov:1977dq}. Conventionally, moments from
the low-energy expansion have been used for this
purpose~\cite{Kuhn:2001dm,Boughezal:2006px,Kuhn:2007vp,Chetyrkin:2009fv,Chetyrkin:2010ic,Dehnadi:2011gc},
but also methods relying on  high-energy moments have been shown to yield
quite accurate results~\cite{Bodenstein:2010qx}.

As an alternative to experimental data also lattice simulations can be
used in order to compute moments of two-point functions. Thanks to significant improvements in the last years
results can be obtained not only for the charm-quark mass and the strong
coupling constant~\cite{Allison:2008xk}, but also for the bottom-quark
mass~\cite{McNeile:2010ji}.

Aside from these determinations, quark-current correlators have a number
of additional phenomenological applications. Two examples are the
sum-rule determinations of heavy-meson decay
constants and $\alpha_s(m_\tau)$ (for
recent works see e.g. Refs ~\cite{Penin:2001ux,Lucha:2010ea,Baikov:2008jh,Davier:2008sk}).  While the
former requires -- among other ingredients -- the low-energy expansion of the
respective correlator, the latter uses the absorptive part above the
production threshold.

In this work we consider correlators of quark currents with various
different Lorentz structures. Furthermore we distinguish between
diagonal and non-diagonal correlators. The diagonal
correlators are defined through
\begin{equation}
\label{eq:4}
\Pi_\mathrm{D}^{X}(q) = i \int e^{i q x } \langle 0| J^X(x) J^X(0)  |0
\rangle \mathrm {d} x   \, ,
\end{equation}
with the currents $J^X(x)$ given by 
\begin{equation}
  \label{eq:3}
J^X(x) = \bar \psi(x) \Gamma^X \psi(x)   \, .
\end{equation}
The non-diagonal correlators are defined in analogy by 
\begin{equation}
  \label{eq:5}
\Pi_\mathrm{ND}^{X}(q) = i \int e^{i q x }\langle 0| j^X(x) j^X(0)  |0
\rangle \mathrm {d} x    \, ,
\end{equation}
with the currents $j^X(x)$ given by 
\begin{equation}
  \label{eq:6}
j^X(x) = \bar \psi(x) \Gamma^X \chi(x)\,.
\end{equation}
Here $\psi$ denotes a heavy quark while $\chi$ stands for a
massless one. The scalar, pseudo-scalar, vector and axial-vector
currents are defined through the Dirac structures $\Gamma^X = 1, i \gamma_5, \gamma_\mu, \gamma_\mu
\gamma_5$ ($X=s,p,v,a$), respectively.  Note that in the case of the non-diagonal
correlators $\Pi_\mathrm{ND}^{s}=\Pi_\mathrm{ND}^{p}$ and  $\Pi_\mathrm{ND}^{v}=\Pi_\mathrm{ND}^{a}$ hold.

For the determination of the quark masses from experimental data only the
diagonal vector correlator $\Pi_\mathrm{D}^v$ can be used while it is
in principle possible to use all diagonal and non-diagonal correlators in
combination with lattice simulations.

Current-current correlators have already been extensively studied in the
li\-terature. The two-loop corrections to the diagonal vector correlator
have been calculated in~\cite{Kallen:1955fb}, the corresponding
corrections for the non-diagonal one in~\cite{Reinders:1981sy}. At three
loops the leading terms of the low- and high-energy expansion have been
determined in~\cite{Chetyrkin:1997mb,Chetyrkin:2001je}. For the diagonal
correlators many terms in the low-energy expansion have been computed
in~\cite{Boughezal:2006uu,Maier:2007yn}. At four loops the leading terms
in the low-energy expansion of various currents have been calculated 
in~\cite{Chetyrkin:2006xg,Sturm:2008eb,Boughezal:2006px,Maier:2008he,Maier:2009fz}. In
the high-energy region the expansion at four-loop order has been
obtained in~\cite{Baikov:2009uw}. In~\cite{Hoff:2011ge} {effects of a
second massive quark } have been taken into account at three-loop order. 
Collecting the available
information for the correlators at three and four loops, the full
momentum dependence has been approximated using Pad\'e approximations
and Mellin-Barnes inspired 
methods~\cite{Chetyrkin:1996cf,Chetyrkin:1997mb,Kiyo:2009gb,Hoang:2008qy,Greynat:2010kx,Greynat:2011zp}. 
In this paper we complete the calculation of the low- and high-energy
expansion of diagonal and non-diagonal heavy-quark correlators
using an improved method for the calculation.

In Section~\ref{sec:details-calculation} we present the details of the
calculation. In Section~\ref{sec:results} we present explicit results for the
non-diagonal vector correlator in numerical form and discuss the convergence of
the expansions. The analytical results for all correlators can be found
on the web page \\http://www-ttp.particle.uni-karlsruhe.de/Progdata/ttp11/ttp11-25/.
\section{Details of the calculation}
\label{sec:details-calculation}
\begin{figure}
  \centering
 \null\hfill \includegraphics[width=0.15\linewidth]{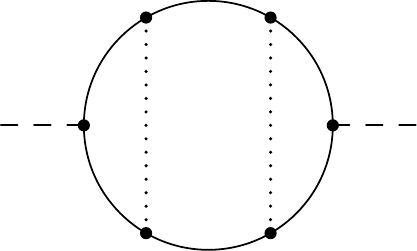}\hfill
  \includegraphics[width=0.15\linewidth]{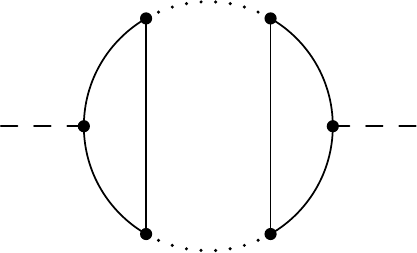}\hfill
  \includegraphics[width=0.15\linewidth]{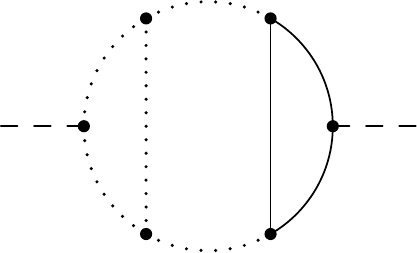}\hfill
  \includegraphics[width=0.15\linewidth]{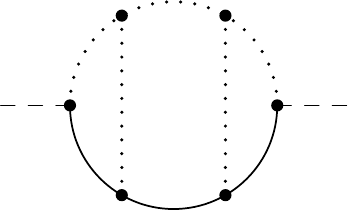}\hfill
\null\\
  \includegraphics[width=0.15\linewidth]{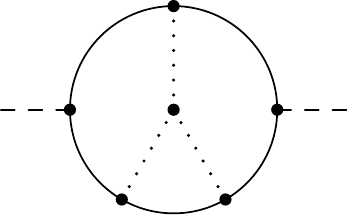}\hfill
  \includegraphics[width=0.15\linewidth]{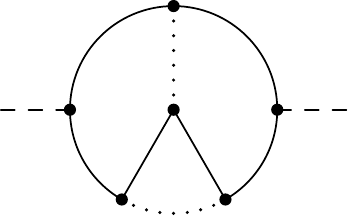}\hfill
  \includegraphics[width=0.15\linewidth]{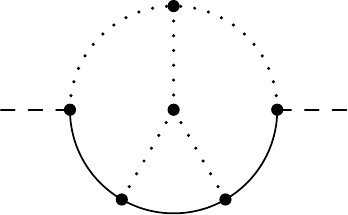}\hfill
  \includegraphics[width=0.15\linewidth]{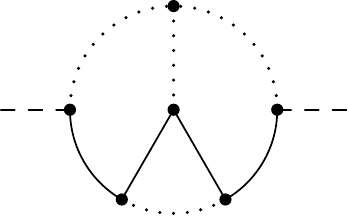}\hfill
  \includegraphics[width=0.15\linewidth]{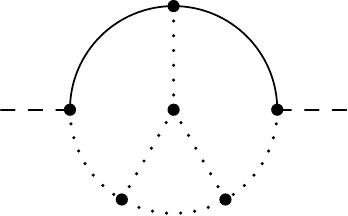}\hfill
\\\null\hfill
  \includegraphics[width=0.15\linewidth]{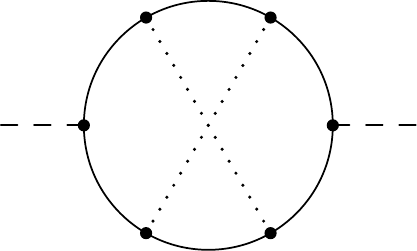}\hfill
  \includegraphics[width=0.15\linewidth]{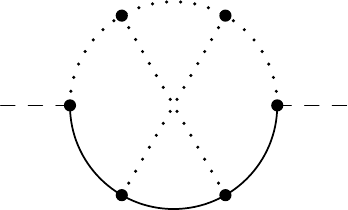}\hfill\null
  \caption{Three-loop master topologies needed for the calculation of
    the non-diagonal and diagonal
    correlators. Solid lines denote massive propagators while dotted
    lines represent massless ones.}
  \label{fig:masterTopos}
\end{figure}

\begin{figure}
  \centering
  \includegraphics[width=\textwidth]{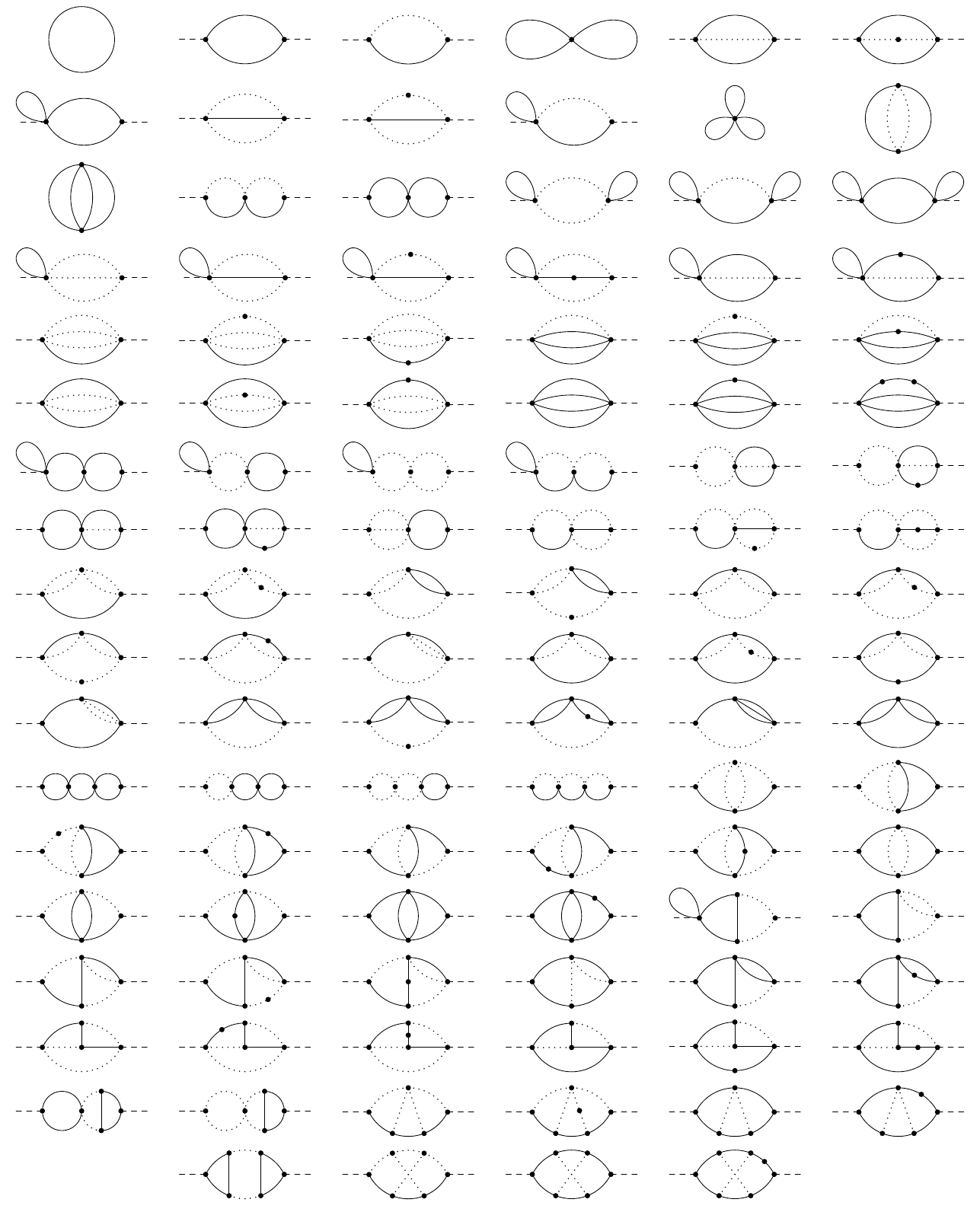}
  \caption{All master integrals needed for the calculation of the
    diagonal and non-diagonal correlators up to three loops. Solid lines denote
    massive propagators while dotted lines represent massless ones. A
    dot on a line denotes a squared propagator.}
  \label{fig:masters}
\end{figure}

The calculation is organized as follows. The diagrams are generated
using {\tt qgraf}~\cite{Nogueira:1991ex} and are subsequently mapped
onto a small set of 11 topologies (cf. Fig. \ref{fig:masterTopos}) using
{\tt q2e} and {\tt exp}~\cite{exp}. After the application of suitable
projectors the resulting scalar integrals are reduced to the master
integrals shown in Fig. \ref{fig:masters} 
using integration-by-parts
identities~\cite{Chetyrkin:1981qh,Laporta:2001dd} implemented in the
program {\tt Crusher}~\cite{crusher}. The resulting expression which
still contains the full dependence on both the external momentum square
$q^2$ and the heavy-quark mass $m$ is then -- lacking an analytical
result for the master integrals -- expanded in the energy region of
interest as explained in more detail below. The expanded master
integrals are combined in a {\tt FORM}~\cite{Vermaseren:2000nd} program
in order to obtain the final expressions valid in either the low-energy
or the high-energy region. For further use the low- and high-energy
expansions of all master integrals can be obtained from the web site http://www-ttp.particle.uni-karlsruhe.de/Progdata/ttp11/ttp11-25/masters/.


The coupling constant
$\alpha_s$ is renormalized in the \MSbar{} scheme while
for the renormalization of the heavy-quark mass $m$ we employ both the
\MSbar{} and the on-shell scheme. In the following we
concentrate on the renormalization in the \MSbar{} scheme, the results for the on-shell scheme can be found on
the web page mentioned above. For the overall normalization we employ $\Pi^X(0)=0$ with $X \in {v,a,s,p}$.

\begin{figure}
  \centering
  \includegraphics[width=.12\linewidth]{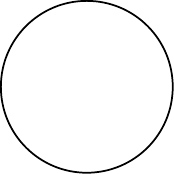}\quad
  \includegraphics[width=.12\linewidth]{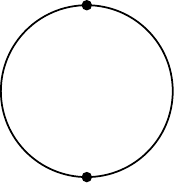}\quad
  \includegraphics[width=.12\linewidth]{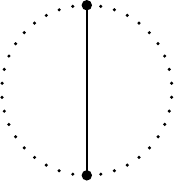}\quad
  \includegraphics[width=.12\linewidth]{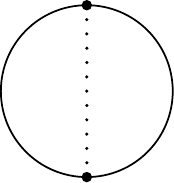}\\[.5em]
  \includegraphics[width=.12\linewidth]{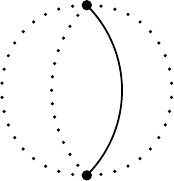}\quad
  \includegraphics[width=.12\linewidth]{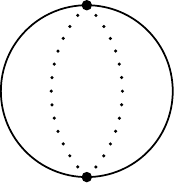}\quad
  \includegraphics[width=.12\linewidth]{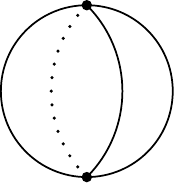}\quad
  \includegraphics[width=.12\linewidth]{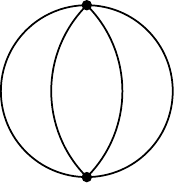}\quad
  \includegraphics[width=.12\linewidth]{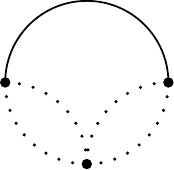}\quad
  \includegraphics[width=.12\linewidth]{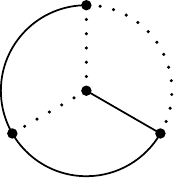}\\[.5em]
\includegraphics[width=.2\linewidth]{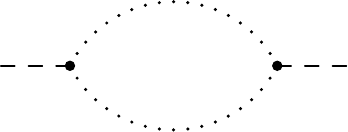}\quad
\includegraphics[width=.2\linewidth]{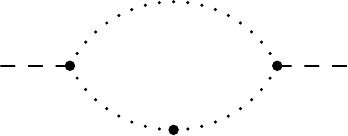}\quad
\includegraphics[width=.2\linewidth]{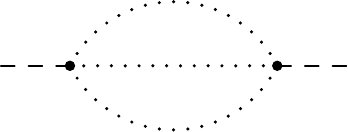}\\[.5em]
\includegraphics[width=.2\linewidth]{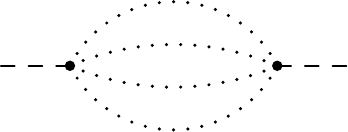}\quad
\includegraphics[width=.2\linewidth]{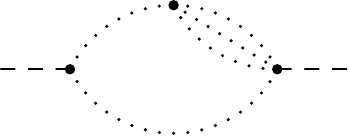}\quad
\includegraphics[width=.2\linewidth]{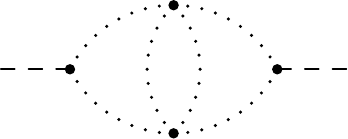}\quad
\includegraphics[width=.2\linewidth]{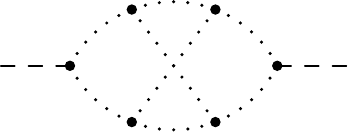}\\[.5em]
  \caption{Building blocks for the initial conditions
    $A_{i,k_0},\,A_{i,k_0}^{(l\epsilon)},\,l=1,2,3$. The initial
    conditions are linear combinations of products of the
    diagrams shown here. The diagrams are evaluated for $-q^2=m^2=1$ in
    order to arrive at dimensionless coefficients
    $A_{i,k_0},\,A_{i,k_0}^{(l\epsilon)}$.} 
 \label{fig:1}
\end{figure}

Let us now have a closer look at the technical details of the
expansion of the master integrals. The starting point for the calculation of
the expansion is a suitable differential equation,
which can easily be obtained from the master formula
\cite{Kotikov:1990kg,Remiddi:1997ny,Caffo:1998du} 
\begin{equation}
  \label{eq:7}
{\cal D}_i M_i (q^2,m^2) = 2 \left(q^2 \frac{\partial}{\partial q^2} + m^2
\frac{\partial}{\partial m^2}\right) M_i(q^2,m^2) ,  
\end{equation}
where ${\cal D}_i$ is the mass dimension of the integral $M_i
(q^2,m^2)$. The derivative with respect to $m^2$ can be performed resulting in
a differential equation with respect to $q^2$ for the integral $M_i
(q^2,m^2)$. The expansion in $z=q^2/m^2$ can now be obtained by inserting a
suitable ansatz, e.g.  
\begin{equation}
 M_i(q^2,m^2) = m^{{\cal D}_i} \sum_{k=0}^\infty A_{i,k} z^k 
\label{eq:1}
\end{equation}
into the differential equation and solving the resulting linear system
of equations for the coefficients $A_{i,k}$. Ansatz (\ref{eq:1}) is
sufficient to obtain the low-energy expansion for the non-diagonal
correlators. The only further input needed is the value of the integral
for $z=0$, which corresponds to the initial condition of the
differential equation. In the case of the high-energy expansion ansatz
(\ref{eq:1}) is not sufficient\footnote{An extension is also necessary
  for the low-energy expansion of the diagonal correlators as discussed in
  detail in Ref.~\cite{Maier:2007yn}.} but one has to use an ansatz of the
form
\begin{equation}
  \label{eq:2}
  M_i(q^2,m^2)=m^{{\cal D}_i} 
  \sum_{k=k_0}^{\infty} \big( A_{i,k} 
  + A_{i,k}^{(\epsilon)} (-z)^{-\epsilon} 
  + A_{i,k}^{(2\epsilon)} (-z)^{-2\epsilon} 
  + A_{i,k}^{(3\epsilon)} (-z)^{-3\epsilon} 
  \big) z^{-k}
\end{equation}
instead. 
The lower limit $k_0$ and the initial conditions
$A_{i,k_0},\,A_{i,k_0}^{(l\epsilon)},\,l=1,2,3$ can be determined
from the leading terms of the asymptotic expansion for $z\to-\infty$.
Generally, these leading terms consist of products of massive vacuum
diagrams and massless propagators with a total number of three loops in
each product. The coefficients $A_{i,k_0}^{(l\epsilon)}$ are given by
the products of propagators with a total of $l$ loops and vacuum
diagrams with $3-l$ loops. Consequently, $A_{i,k_0}$ consists of the
products without any massless propagators. The diagrams contributing to
the initial conditions are shown in Fig. \ref{fig:1}; in order to obtain
dimensionless coefficients $A_{i,k_0},\,A_{i,k_0}^{(l\epsilon)}$ we set
$-q^2=m^2=1$ in those diagrams.  

It should be noted that in practice
it is convenient to just use a small enough
universal value for $k_0$ instead of carefully deriving it separately for
each single master integral from the respective asymptotic
expansion. Furthermore, most of the coefficients
$A_{i,k_0},\,A_{i,k_0}^{(l\epsilon)}$ are already determined by the
linear system of equations and only a few
remaining ones have to be determined through asymptotic expansion. The
considerable freedom in the choice of the initial conditions also
implies that the set of contributing diagrams shown in Fig. \ref{fig:1} is by no
means unique.  

Finally, let us show as an
example the explicit high-energy expansion of one specific master
integral:
\begin{align}
  M_{61}(q^2,m^2)=&\,\mathgraphics{80pt}{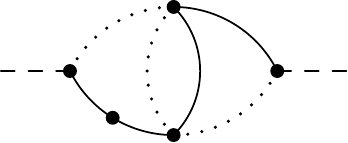}
\,,\notag\\
    \sum A_{k}z^{-k}=&\,
\big ( z^{-3} + \ldots \big)
\,\mathgraphics{30pt}{T1L11}^3+\big(3\*z^{-3}+z^{-2}+ \ldots \big)\,\mathgraphics{30pt}{T3L42} \,, \notag\\
    \sum
    A_{k}^{(\epsilon)}z^{-k}=&\big(-6\*z^{-3} + \ldots \big ) \*\,\mathgraphics{50pt}{P1L21}\
    \mathgraphics{30pt}{T1L11}^2 \,,\notag\\
\sum
A_{k}^{(2\epsilon)}z^{-k}=&\big(-8\*\epsilon^{-1}\*z^{-3}+12\*\epsilon^{-1}\*z^{-2}-2\*\epsilon^{-1}\*z^{-1}+99\*z^{-3}\notag\\
&\qquad-48\*z^{-2}+5\*z^{-1} + \ldots\big)\,\mathgraphics{50pt}{P2L31}\*\ \mathgraphics{30pt}{T1L11}\notag\\
&+\big(4\*z^{-3}+2\*z^{-2}+\ldots\big)\,\mathgraphics{50pt}{P1L21}^2\,\mathgraphics{30pt}{T1L11}\,,\notag\\
\sum
A_{k}^{(3\epsilon)}z^{-k}=&\big(-36\*\epsilon^{-2}\*z^{-3}+36\*\epsilon^{-2}\*z^{-2}-6\*\epsilon^{-2}\*z^{-1}+564\*\epsilon^{-1}\*z^{-3}-294\*\epsilon^{-1}\*z^{-2}\notag\\
&+29\*\epsilon^{-1}\*z^{-1}-2665\*z^{-3}+508\*z^{-2}-46\*z^{-1}+\ldots\big)\,\mathgraphics{50pt}{P3L41}\notag\\
&+\big(-3\*z^{-3}-z^{-2}-z^{-1}+\ldots\big)\*\,\mathgraphics{50pt}{P3L61}\,.
  \label{eq:8}
\end{align} 
 In this example terms of order $\epsilon$ or $1/z^4$ and higher have been
 omitted in the coefficients of the integrals.

In the case of the pseudo-scalar and axial-vector correlators one has to address
the treatment of $\gamma_5$. In this calculation we use the
prescription given by Larin~\cite{Larin:1993tq} for the so-called singlet diagrams with exactly one $\gamma_5$
 in a fermion trace and a na\"ively anticommuting $\gamma_5$ for all
 other diagrams. In order to present only anomaly-free quantities
the axial-vector correlator is computed for a quark isospin
 doublet $(\psi,\chi)$ like in Ref.~\cite{Maier:2007yn}.

\section{Results}
\label{sec:results}
Since the full results are very lengthy we present only the numerical
results for the non-diagonal vector correlator. The analytical results
for all correlators can be found on the web page\\
http://www-ttp.particle.uni-karlsruhe.de/Progdata/ttp11/ttp11-25/.

The expansion has been performed in $z=q^2 /m^2$. The results for the
first 6 low-energy and 8 high-energy coefficients agree with
Ref.~\cite{Chetyrkin:2001je}.

The vector correlator can be written as 
\begin{equation}
  \label{eq:9}
\Pi_{\mathrm{ND},v}^{\mu\nu}(q^2) =   (- q^2 g^{\mu\nu} +
q^\mu q^\nu)\Pi_{\mathrm{ND}}^{v}(q^2) + q^\mu q^\nu \Pi^v_{\mathrm{
    ND,L}} \, ,
\end{equation}
{ We present only the results for the transversal part of the vector
  correlator, the longitudinal part can be obtained from the results for
  the scalar one. In the following we set the renormalization scale $\mu=m$
  during the numerical evaluation}.

In Table~\ref{tab:nd_v_MS_le_3l} we present the numerical
results for the low-energy expansion of the non-diagonal vector correlator
which is decomposed according to
\begin{equation}
\begin{split}
  \label{eq:5a}
\Pi_\mathrm{ND}^{v} =& \frac{3}{16 \pi^2}\sum_{n>0} \bigg(
    \bar{C}^{(0),v}_n 
+  \frac{\alpha_s}{\pi}
  \bar{C}^{(1),v}_n  \\&+  \left ( \frac{\alpha_s}{\pi} \right )^2\left (  \bar{C}^{(2),v}_n[1]
  + n_h \bar{C}^{(2),v}_n[n_h] + n_l \bar{C}^{(2),v}_n[n_l]\right ) \bigg) z^n   \, .
\end{split}
\end{equation}
$\bar{C}^{(2),v}_n[n_h]$ and $\bar{C}^{(2),v}_n[n_l]$ contain the
contributions from closed heavy and light quark loops, respectively. 

The results for the high-energy expansion shown in Tables 2--4 are more
complicated due to logarithmic contributions. Since the correlator
is known analytically up to two loops we present only the results for the
three-loop contribution. It is decomposed as
\begin{equation}
\begin{split}
  \label{eq:5b}
  \Pi_\mathrm{ND}^{v,\mathrm{3-loop}} =& 
\frac{3}{16 \pi^2}\left ( \frac{\alpha_s}{\pi} \right )^2
\\&\times 
\sum_{n\geq0} \sum_{m=0}^3\Big (  
  \bar{D}^{(2),v}_{n,m}[1] 
  + n_h \bar{D}^{(2),v}_{n,m}[n_h]   
  + n_l \bar{D}^{(2),v}_{n,m}[n_l] \Big)
{z^{-n}}\ln^m{\frac{-q^2}{m^2}}    \,.
\end{split}
\end{equation}

\begin{figure}
  \centering
  \includegraphics[width=0.7\textwidth]{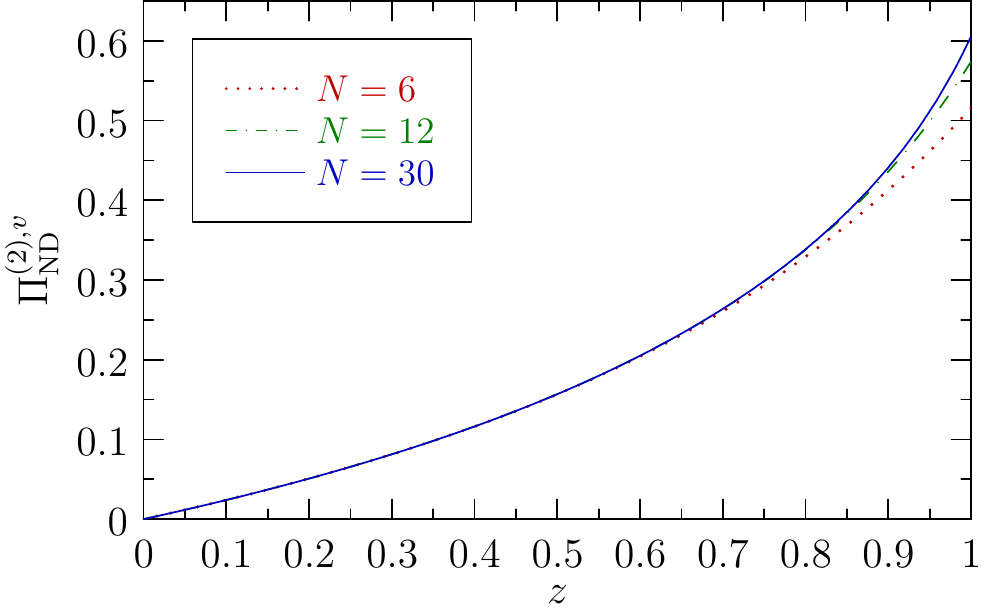}
  \includegraphics[width=0.7\textwidth]{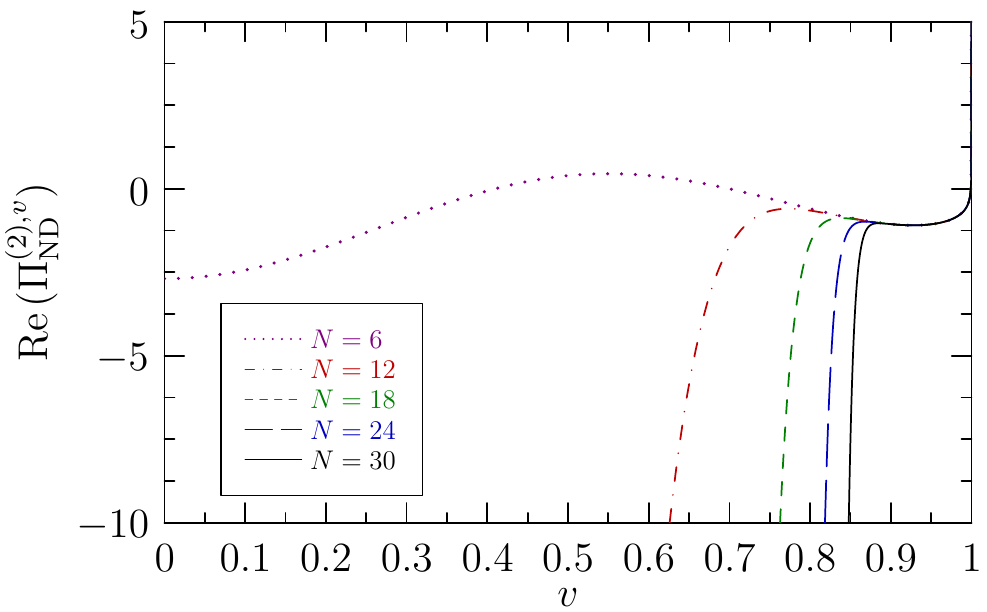}
  \includegraphics[width=0.7\textwidth]{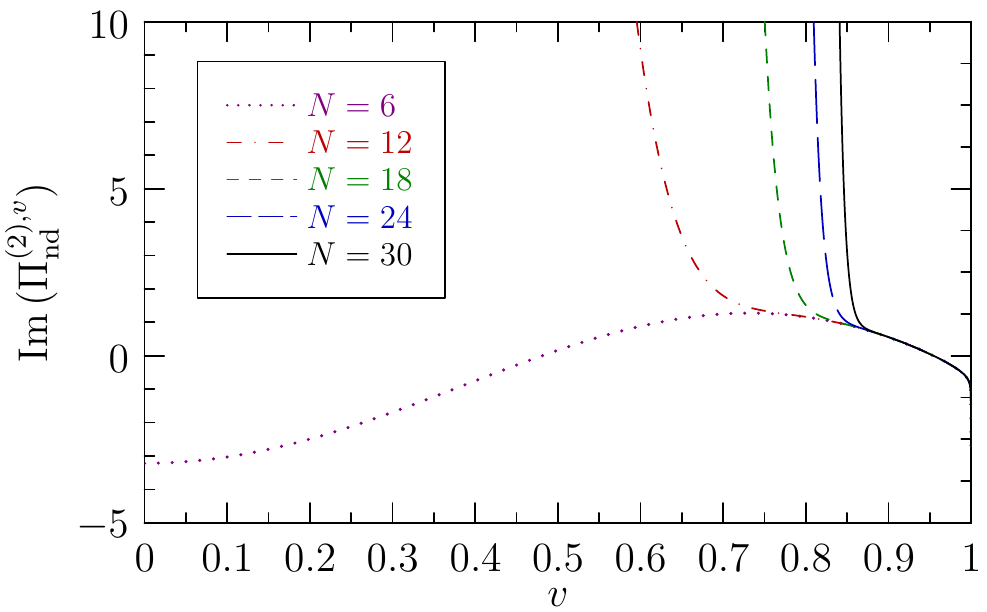}
  \caption{Convergence of the three-loop contribution to the low- and
    high-energy expansion of the non-diagonal vector correlator in the
    \MSbar{} scheme for $n_h=1$, $n_l=3$. The upper figure shows the
    correlator in the low-energy range for different orders of the
    expansion in $z$. The lower figures show the behavior above
    threshold for both real and imaginary parts. Note that the
    high-energy expansion converges only above $v=0.8$ which
    corresponds to the three-particle cut.}
  \label{fig:nd_vec_os}
\end{figure}

In Fig.~\ref{fig:nd_vec_os} we show the convergence of the series for
the case of the non-diagonal vector correlator. The behavior of the
series is shown both for the low-energy and the high-energy
expansion. For the high-energy expansion we show the real and the imaginary
part. Since the non-diagonal correlators have a cut through three heavy
quark lines, the high-energy expansion converges only above $z=9$, which
corresponds to a velocity of the heavy quark of
$v=(z-1)/(z+1)=0.8$.\footnote{The diagonal correlators have a
  four-particle cut starting at three-loops and converge therefore only
  above $z=16$, i.e. $v=\sqrt{1-1/z}=0.97$.} This behavior can also clearly be observed in
Tables 2 and 3 which show rapidly growing coefficients corresponding to
a radius of convergence $r<1$. Since Table 4 contains only contributions
from diagrams with a closed massless-quark loop, the coefficients are
better behaved. 
\begin{table}
\renewcommand{\arraystretch}{\pmstretch}
  \centering
\begin{tabular}{rlllll}
\hline
$n$ & $\bar{C}^{(0),v}_n$ & $\bar{C}^{(1),v}_n$ & $\bar{C}^{(2),v}_n[1]$
&$\bar{C}^{(2),v}_n[n_h]$ &$\bar{C}^{(2),v}_n[n_l]$\\\hline
$1$&$0.500000$&$0.457734$&$-1.51195$&$0.0166564$&$0.337837 $\\
$2$&$0.133333$&$-0.0483080$&$-0.840974$&$-0.0369934$&$0.187255 $\\
$3$&$0.0555556$&$-0.115602$&$-0.665554$&$-0.0289884$&$0.135615 $\\
$4$&$0.0285714$&$-0.115219$&$-0.475376$&$-0.0213752$&$0.104474 $\\
$5$&$0.0166667$&$-0.102152$&$-0.309702$&$-0.0161172$&$0.0832408 $\\
$6$&$0.0105820$&$-0.0880811$&$-0.177592$&$-0.0125094$&$0.0679630 $\\
$7$&$0.00714286$&$-0.0756354$&$-0.0753998$&$-0.00996452$&$0.0565728 $\\
$8$&$0.00505051$&$-0.0651894$&$0.00291844$&$-0.00811379$&$0.0478459 $\\
$9$&$0.00370370$&$-0.0565463$&$0.0628596$&$-0.00672995$&$0.0410083 $\\
$10$&$0.00279720$&$-0.0494003$&$0.108797$&$-0.00566985$&$0.0355496 $\\
$11$&$0.00216450$&$-0.0434648$&$0.144058$&$-0.00484063$&$0.0311210 $\\
$12$&$0.00170940$&$-0.0385013$&$0.171144$&$-0.00418016$&$0.0274778 $\\
$13$&$0.00137363$&$-0.0343196$&$0.191926$&$-0.00364579$&$0.0244440 $\\
$14$&$0.00112045$&$-0.0307699$&$0.207816$&$-0.00320745$&$0.0218902 $\\
$15$&$0.000925926$&$-0.0277347$&$0.219882$&$-0.00284350$&$0.0197199 $\\
$16$&$0.000773994$&$-0.0251214$&$0.228940$&$-0.00253805$&$0.0178596 $\\
$17$&$0.000653595$&$-0.0228568$&$0.235620$&$-0.00227923$&$0.0162527 $\\
$18$&$0.000556948$&$-0.0208824$&$0.240408$&$-0.00205802$&$0.0148551 $\\
$19$&$0.000478469$&$-0.0191512$&$0.243686$&$-0.00186749$&$0.0136316 $\\
$20$&$0.000414079$&$-0.0176253$&$0.245754$&$-0.00170221$&$0.0125545 $\\
$21$&$0.000360750$&$-0.0162738$&$0.246850$&$-0.00155793$&$0.0116011 $\\
$22$&$0.000316206$&$-0.0150713$&$0.247164$&$-0.00143123$&$0.0107532 $\\
$23$&$0.000278707$&$-0.0139968$&$0.246847$&$-0.00131937$&$0.00999562 $\\
$24$&$0.000246914$&$-0.0130328$&$0.246023$&$-0.00122012$&$0.00931596 $\\
$25$&$0.000219780$&$-0.0121648$&$0.244790$&$-0.00113166$&$0.00870382 $\\
$26$&$0.000196483$&$-0.0113806$&$0.243229$&$-0.00105247$&$0.00815053 $\\
$27$&$0.000176367$&$-0.0106696$&$0.241406$&$-0.000981314$&$0.00764873 $\\
$28$&$0.000158907$&$-0.0100231$&$0.239376$&$-0.000917133$&$0.00719220 $\\
$29$&$0.000143678$&$-0.00943357$&$0.237182$&$-0.000859046$&$0.00677564 $\\
$30$&$0.000130336$&$-0.00889448$&$0.234861$&$-0.000806306$&$0.00639448
$\\\hline
\end{tabular}
  \caption{Numerical results for the low-energy moments for the non-diagonal vector correlator in
    the \MSbar{} scheme.
}
  \label{tab:nd_v_MS_le_3l}
\end{table}
\begin{table}
\renewcommand{\arraystretch}{\pmstretch}
  \centering
  \begin{tabular}{lcccc}
\hline
  $n$& $\bar{D}^{(2),v}_{n,0}[1]$& $\bar{D}^{(2),v}_{n,1}[1]$&
  $\bar{D}^{(2),v}_{n,2}[1]$& $\bar{D}^{(2),v}_{n,3}[1]$\\\hline
$0$&$1.78284$&$-2.64761$&$1.83333$&$0$ \\
$1$&$-17.7072$&$25.1337$&$-16.3333$&$3.16667$ \\
$2$&$-60.1359$&$47.0000$&$-8.00000$&$0$ \\
$3$&$-41.3176$&$-15.4431$&$16.0772$&$-3.32442$ \\
$4$&$4.27239$&$-20.8764$&$5.72599$&$0.659122$ \\
$5$&$7.18683$&$-11.8251$&$0.822323$&$-0.129739$ \\
$6$&$0.493132$&$-5.22086$&$2.92024$&$-0.566886$ \\
$7$&$-16.8712$&$-10.5240$&$9.71602$&$-0.991008$ \\
$8$&$-56.8660$&$-43.6415$&$29.4833$&$-1.70280$ \\
$9$&$-164.387$&$-190.388$&$94.9042$&$-2.87364$ \\
$10$&$-485.549$&$-831.419$&$339.294$&$-4.95108$ \\
$11$&$-1537.17$&$-3780.76$&$1357.43$&$-8.63056$ \\
$12$&$-5240.01$&$-18141.8$&$6001.46$&$-15.2756$ \\
$13$&$-18953.8$&$-91894.0$&$28790.4$&$-27.3450$ \\
$14$&$-70903.6$&$-488762$&$147374$&$-49.4690$ \\
$15$&$-264436$&$-2.71202\cdot10^6$&$794591$&$-90.2629$ \\
$16$&$-922552$&$-1.56052\cdot10^7$&$4.46970\cdot10^6$&$-165.935$ \\
$17$&$-2.52510\cdot10^6$&$-9.26482\cdot10^7$&$2.60485\cdot10^7$&$-306.988$ \\
$18$&$-142055$&$-5.65199\cdot10^8$&$1.56448\cdot10^8$&$-571.093$ \\
$19$&$8.11996\cdot10^7$&$-3.53100\cdot10^9$&$9.64408\cdot10^8$&$-1067.52$ \\
$20$&$9.74537\cdot10^8$&$-2.25279\cdot10^{10}$&$6.08196\cdot10^9$&$-2003.91$ \\
$21$&$8.94750\cdot10^9$&$-1.46443\cdot10^{11}$&$3.91351\cdot10^{10}$&$-3775.71$ \\
$22$&$7.44796\cdot10^{10}$&$-9.68039\cdot10^{11}$&$2.56374\cdot10^{11}$&$-7137.78$ \\
$23$&$5.91860\cdot10^{11}$&$-6.49650\cdot10^{12}$&$1.70673\cdot10^{12}$&$-13533.8$ \\
$24$&$4.59022\cdot10^{12}$&$-4.41990\cdot10^{13}$&$1.15282\cdot10^{13}$&$-25729.9$ \\
$25$&$3.51344\cdot10^{13}$&$-3.04480\cdot10^{14}$&$7.88989\cdot10^{13}$&$-49035.5$ \\
$26$&$2.67074\cdot10^{14}$&$-2.12156\cdot10^{15}$&$5.46505\cdot10^{14}$&$-93657.0$ \\
$27$&$2.02372\cdot10^{15}$&$-1.49382\cdot10^{16}$&$3.82727\cdot10^{15}$&$-179244$ \\
$28$&$1.53213\cdot10^{16}$&$-1.06201\cdot10^{17}$&$2.70750\cdot10^{16}$&$-343676$ \\
$29$&$1.16067\cdot10^{17}$&$-7.61776\cdot10^{17}$&$1.93326\cdot10^{17}$&$-660072$ \\
$30$&$8.80664\cdot10^{17}$&$-5.50954\cdot10^{18}$&$1.39237\cdot10^{18}$&$-1.26974\cdot10^6
$\\\hline
\end{tabular}
\caption{Three-loop coefficients of the high-energy expansion for the non-diagonal
  vector correlator in the \MSbar{} scheme with $n_h=n_l=0$.}
\label{tab:nd_v_he_quenched}
\end{table}
\begin{table}
\renewcommand{\arraystretch}{\pmstretch}
  \centering
  \begin{tabular}{lcccc}
\hline
$n$&$\bar{D}^{(2),v}_{n,0}[n_h]$ &$\bar{D}^{(2),v}_{n,1}[n_h]$
&$\bar{D}^{(2),v}_{n,2}[n_h]$ &$\bar{D}^{(2),v}_{n,3}[n_h]$ \\\hline
$0$&$-0.251201$&$0.153727$&$-0.111111$&$0$ \\
$1$&$0.154182$&$-1.14726$&$0.555556$&$-0.111111$ \\
$2$&$3.61746$&$-1.14459$&$-0.666667$&$0$ \\
$3$&$1.27971$&$2.67597$&$1.06790$&$0.185185$ \\
$4$&$-3.99022$&$-1.81636$&$-0.861111$&$-0.543210$ \\
$5$&$-8.59452$&$0.842895$&$1.23685$&$-1.15062$ \\
$6$&$-25.2903$&$-2.26222$&$8.70815$&$-1.68560$ \\
$7$&$-65.3844$&$-31.4942$&$28.2131$&$-2.18624$ \\
$8$&$-164.223$&$-160.905$&$86.7525$&$-2.66878$ \\
$9$&$-441.129$&$-715.641$&$295.743$&$-3.14070$ \\
$10$&$-1320.45$&$-3233.48$&$1154.83$&$-3.60588$ \\
$11$&$-4384.76$&$-15440.6$&$5072.08$&$-4.06652$ \\
$12$&$-15690.9$&$-78173.7$&$24362.3$&$-4.52396$ \\
$13$&$-58311.1$&$-416730$&$125176$&$-4.97906$ \\
$14$&$-215239$&$-2.32045\cdot10^6$&$677824$&$-5.43239$ \\
$15$&$-732644$&$-1.34050\cdot10^7$&$3.82947\cdot10^6$&$-5.88438$ \\
$16$&$-1.83932\cdot10^6$&$-7.99084\cdot10^7$&$2.24126\cdot10^7$&$-6.33529$ \\
$17$&$1.84847\cdot10^6$&$-4.89431\cdot10^8$&$1.35168\cdot10^8$&$-6.78534$ \\
$18$&$8.23216\cdot10^7$&$-3.06952\cdot10^9$&$8.36569\cdot10^8$&$-7.23470$ \\
$19$&$9.21471\cdot10^8$&$-1.96567\cdot10^{10}$&$5.29604\cdot10^9$&$-7.68348$ \\
$20$&$8.27344\cdot10^9$&$-1.28233\cdot10^{11}$&$3.42033\cdot10^{10}$&$-8.13178$ \\
$21$&$6.81847\cdot10^{10}$&$-8.50550\cdot10^{11}$&$2.24852\cdot10^{11}$&$-8.57968$ \\
$22$&$5.39132\cdot10^{11}$&$-5.72647\cdot10^{12}$&$1.50188\cdot10^{12}$&$-9.02723$ \\
$23$&$4.17068\cdot10^{12}$&$-3.90796\cdot10^{13}$&$1.01766\cdot10^{13}$&$-9.47449$ \\
$24$&$3.18860\cdot10^{13}$&$-2.69998\cdot10^{14}$&$6.98579\cdot10^{13}$&$-9.92149$ \\
$25$&$2.42300\cdot10^{14}$&$-1.88650\cdot10^{15}$&$4.85260\cdot10^{14}$&$-10.3683$ \\
$26$&$1.83633\cdot10^{15}$&$-1.33179\cdot10^{16}$&$3.40755\cdot10^{15}$&$-10.8149$ \\
$27$&$1.39098\cdot10^{16}$&$-9.49177\cdot10^{16}$&$2.41677\cdot10^{16}$&$-11.2613$ \\
$28$&$1.05453\cdot10^{17}$&$-6.82454\cdot10^{17}$&$1.72987\cdot10^{17}$&$-11.7075$ \\
$29$&$8.00841\cdot10^{17}$&$-4.94695\cdot10^{18}$&$1.24877\cdot10^{18}$&$-12.1537$ \\
$30$&$6.09583\cdot10^{18}$&$-3.61318\cdot10^{19}$&$9.08598\cdot10^{18}$&$-12.5997
$\\\hline
\end{tabular}
\caption{Three-loop contribution proportional to $n_h$ to the
  coefficients of the high-energy expansion for the non-diagonal 
  vector correlator in the \MSbar{} scheme.}
\label{tab:nd_v_he_nh}
\end{table}
\begin{table}
\renewcommand{\arraystretch}{\pmstretch}
  \centering
  \begin{tabular}{lcccc}
\hline
$n$&$\bar{D}^{(2),v}_{n,0}[n_l]$ &$\bar{D}^{(2),v}_{n,1}[n_l]$
&$\bar{D}^{(2),v}_{n,2}[n_l]$ &$\bar{D}^{(2),v}_{n,3}[n_l]$\\\hline
$0$&$-0.730175$&$0.153727$&$-0.111111$&$0$ \\
$1$&$-0.314922$&$-1.14726$&$0.555556$&$-0.111111$ \\
$2$&$0.984271$&$-1.77778$&$0$&$0$ \\
$3$&$1.27530$&$-0.0592818$&$-0.615226$&$0.0534979$ \\
$4$&$0.544450$&$0.835905$&$-0.0709877$&$-0.0246914$ \\
$5$&$0.0510031$&$0.410862$&$0.0306790$&$-0.00740741$ \\
$6$&$-0.0437552$&$0.209240$&$0.0303704$&$-0.00329218$ \\
$7$&$-0.0561161$&$0.117605$&$0.0233371$&$-0.00176367$ \\
$8$&$-0.0503344$&$0.0722787$&$0.0177016$&$-0.00105820$ \\
$9$&$-0.0420848$&$0.0475365$&$0.0136697$&$-0.000685871$ \\
$10$&$-0.0347091$&$0.0329283$&$0.0107911$&$-0.000470312$ \\
$11$&$-0.0287111$&$0.0237565$&$0.00869670$&$-0.000336700$ \\
$12$&$-0.0239493$&$0.0177088$&$0.00713792$&$-0.000249408$ \\
$13$&$-0.0201760$&$0.0135592$&$0.00595223$&$-0.000189934$ \\
$14$&$-0.0171671$&$0.0106166$&$0.00503229$&$-0.000148000$ \\
$15$&$-0.0147456$&$0.00847129$&$0.00430581$&$-0.000117578$ \\
$16$&$-0.0127769$&$0.00687002$&$0.00372304$&$-0.0000949668$ \\
$17$&$-0.0111601$&$0.00565028$&$0.00324902$&$-0.0000778089$ \\
$18$&$-0.00981938$&$0.00470458$&$0.00285863$&$-0.0000645526$ \\
$19$&$-0.00869741$&$0.00395987$&$0.00253355$&$-0.0000541477$ \\
$20$&$-0.00775052$&$0.00336532$&$0.00226014$&$-0.0000458663$ \\
$21$&$-0.00694509$&$0.00288479$&$0.00202813$&$-0.0000391926$ \\
$22$&$-0.00625501$&$0.00249212$&$0.00182965$&$-0.0000337544$ \\
$23$&$-0.00565976$&$0.00216805$&$0.00165859$&$-0.0000292783$ \\
$24$&$-0.00514313$&$0.00189817$&$0.00151016$&$-0.0000255604$ \\
$25$&$-0.00469214$&$0.00167159$&$0.00138059$&$-0.0000224467$ \\
$26$&$-0.00429635$&$0.00147992$&$0.00126683$&$-0.0000198192$ \\
$27$&$-0.00394727$&$0.00131666$&$0.00116643$&$-0.0000175864$ \\
$28$&$-0.00363797$&$0.00117672$&$0.00107740$&$-0.0000156771$ \\
$29$&$-0.00336274$&$0.00105607$&$0.000998082$&$-0.0000140345$ \\
$30$&$-0.00311683$&$0.000951473$&$0.000927135$&$-0.0000126137
$\\\hline
\end{tabular}
\caption{Three-loop contribution proportional to $n_l$ to the
  coefficients of the high-energy expansion for the non-diagonal
  vector correlator in the \MSbar{} scheme.}
  \label{tab:nd_v_he_nl}
\end{table}
\clearpage
\section{Conclusion}
\label{sec:conclusion}
We have used modern techniques to compute 30 terms in the low-energy
expansion of the non-diagonal scalar and vector correlators, matching
the already available information on diagonal
correlators~\cite{Maier:2007yn}.  Using slightly modified methods we
have also obtained 30 coefficients in the high-energy expansion of both
non-diagonal and diagonal correlators. 
 Due to multiply massive cuts in some of
the contributing diagrams it is expected that the high-energy expansion
already breaks down significantly above the physical threshold.
Using only the previously available coefficients this behaviour is
rather difficult to see. The newly computed coefficients, however, 
clearly support the presence of such a divergence.

As a final remark it should be noted that the expansion via differential
equations used in this publication is rather efficient. This means that
it is easy to obtain significantly more terms in the expansions if the need
arises.
\section*{Acknowledgments}\label{sec:acknowledgements}

This work was supported by the Deutsche Forschungsgemeinschaft through
the SFB/TR-9 ``Computational Particle Physics''. 
A.\,M. thanks the Landesgraduiertenf\"orderung for support. We would like to
thank T.~Kasprzik, J.H.~K\"uhn and M.~Steinhauser for carefully reading
the manuscript.

\clearpage
\bibliographystyle{hep}
\bibliography{hep}

\end{document}